

\magnification=1200
\baselineskip=14pt

\def\cl{\centerline}

\ \

\bigskip\bigskip
\bigskip\bigskip
\bigskip\bigskip

\cl{\bf GROSS SPECTRAL DIFFERENCES BETWEEN }
\cl{\bf BRIGHT AND DIM GAMMA-RAY BURSTS }

\bigskip\bigskip
\bigskip\bigskip

\cl{R.J. Nemiroff$^{(1),(2)}$, J.P. Norris$^{(1)}$, J. T.
Bonnell$^{(1),(3)}$, }
\cl{ W.A.D.T. Wickramasinghe$^{(4)}$, C. Kouveliotou$^{(3),(5)}$, }
\cl{ W.S. Paciesas$^{(5),(6)}$, G.J. Fishman$^{(5)}$, C. A. Meegan$^{(5)}$ }

\bigskip\bigskip
\bigskip\bigskip

\cl{\it $^{(1)}$ NASA Goddard Space Flight Center, Greenbelt, MD 20771 }
\cl{\it $^{(2)}$ George Mason University, CSI Institute, Fairfax, VA 22030 }
\cl{\it $^{(3)}$ Universities Space Research Association }
\cl{\it $^{(4)}$ Dept. Astronomy, University of Pennsylvania, Philadelphia,
PA 19104 }
\cl{\it $^{(5)}$ NASA Marshall Space Flight Center, Huntsville, AL 35812 }
\cl{\it $^{(6)}$ University of Alabama, Huntsville, AL 35899 }

\bigskip\bigskip
\bigskip\bigskip
\bigskip\bigskip
\bigskip\bigskip
\bigskip\bigskip
\bigskip\bigskip

\cl{ In press: The Astrophysical Journal (Letters) }

\vfill\eject
\baselineskip=18pt

\ \
\bigskip\bigskip

\cl{\bf ABSTRACT }

\bigskip

We find that dim gamma-ray bursts (GRBs) are softer than bright GRBs, as
indicated on average by data from the Burst and Transient Source Experiment
(BATSE) on board the Compton Gamma Ray Observatory.  We show that this
correlation is statistically significant with respect to variations due to
random differences between GRBs.  This effect is discernable using a
variety of methods and data sets, including public domain data.  We analyze
several types of systematic errors and selection effects in the BATSE data
and conclude that the observed effect is not dominated by any of them. We
therefore assert that this dim/soft effect is a real property of GRBs.  It
is possible that this correlation is a consequence of the time dilation
detected by Norris et al. (1994) and that the burst sources are located at
cosmological distances.

\bigskip

\noindent {\it Subject headings:} cosmology:  theory - gamma rays:  bursts

\vfill\eject

\cl{\bf 1. INTRODUCTION }

\bigskip

The Burst and Transient Source Experiment (BATSE) on board NASA's orbiting
{\it Compton} Gamma Ray Observatory has yielded more data on gamma-ray
bursts (GRBs) than any other instrument.  The observed isotropic
distribution of bursts (Meegan et al. 1992) combined with a relative
paucity of dim bursts (Fishman et al. 1992) have failed to bolster the
generally accepted Galactic origin of GRBs that had emerged from analysis
of data of brighter bursts from satellites in the 1980s (Atteia et al.
1987). Today, the origin of GRBs still remains an enigma, with more than
100 models published in the refereed literature (Nemiroff 1994). Paczynski
(1992) and Piran (1992) interpreted early BATSE data as consistent with a
cosmological origin of GRBs and predicted that more distant bursts would
show both a time dilation and a spectral shift relative to closer bursts.
The isotropic distribution of bursts detected by earlier spacecraft also
prompted expectations of cosmological effects (Usov \& Chibisov 1975, van
den Berg 1983). Recently, Norris et al. (1994) and Davis et al. (1994) have
reported measurements of a relative time dilation of order two between
bright and dim GRBs.

Previous claims of spectral differences between bright and dim GRBs were
reported by Mitrofanov et al. (1992a, 1992b, 1993) who analyzed data from
62 GRBs from the Soviet-French experiment APEX on board the {\it Phobos 2}
spacecraft.  Paciesas et al. (1992) found a similar effect with analysis of
the first 126 GRBs detected by BATSE.  Here we present new evidence that
dim GRBs, selected from the first 750 GRBs recorded by BATSE, do indeed
show gross spectral differences compared to bright GRBs.  This effect was
initially apparent but only briefly described in Norris et al. (1994),
appearing as a difference in average peak intensities.  In this paper we
quantify the significance of the effect in several ways, including as a
function of time.  Mitrofanov et al. (1994) also report the same effect in
a smaller sample, but characterize the uncertainties in terms of
statistical errors only - they did not evaluate hardness variations in
different peak intensity samples. The correlation of spectral hardness and
peak intensity is distinct from the evidence reported by Dezalay et al.
(1992) and Kouveliotou et al. (1993) that short ($<$ 2 s) duration GRBs
tend to be spectrally harder than long ($>$ 2 s) GRBs. Spectral properties
of bright BATSE detected GRBs are also discussed by Schaefer et al. (1992,
1994) and Band et al. (1993).

In \S 2 we present our evidence for gross spectral differences, comparing
bursts in the same peak intensity ranges in which relative time dilation
was found by Norris et al. (1994). In \S 3 we present evidence that such an
effect is discernable, although less statistically significant, in 482
bursts contained in the second BATSE catalog with recorded fluences. In \S
4 we discuss possible systematic errors and in \S 5 we discuss the results.

\bigskip

\cl{\bf 2. PRIMARY DATA AND RESULTS }

We inspected all GRB data recorded by BATSE before 15 September 1993 and
created two distinct subsets of GRBs based on their wavelet smoothed 256-ms
peak count rate $P$ - a ``bright" subset of GRBS with $18,000 < P <
250,000$ and a ``dim" subset with $1400 < P < 4500$, where $P$ is measured
in counts sec$^{-1}$ in all triggered BATSE Large Area Detectors (LADs) in
all energy channels combined.  Only GRBs with durations greater than 1.5
seconds were included, for reasons discussed in Norris et al. (1994). There
were 45 GRBs in the bright subset, which comprises $\sim$ 6 \% of the GRBs
measured by BATSE at the time, and 114 in the dim subset, which is near the
flux where BATSE is 99 \% complete in detecting GRbs.  From the work of
Wickramasinghe et al. (1993) and consideration of peak fluxes for bursts
near the low end of the peak intensity range of our bright sample, we find
that the brights are potentially at a significantly smaller redshift than
the dim bursts, assuming that GRBs are cosmological.

We define a hardness ratio, $H(t)$, designed to be relatively insusceptible
to coordinate singularities (such as zero appearing in the denominator) and
at the same time not biased toward bright events:
 $$ H(t) = { \sum_{k} 2 \ C_3^k(t) / (C_2^k(t) + C_3^k(t))
       \over \sum_{k} 2 \ C_2^k(t) / (C_2^k(t) + C_3^k(t)) } .
 \eqno(1)$$
Here $t$ designates time relative to the time of peak intensity, $C_i^k(t)$
stands for counts of GRB \#$k$ above background in energy channel $i$,
where channels 2 and 3 are approximately 50 - 100 keV and 100 - 300 keV,
respectively.  When either $C_2^k(t)$ or $C_3^k(t)$ is less than two sigma
above the respective background for that channel for any $k$, the
contribution of that burst $k$ was excluded from the sum in equation (1).

We utilized approximately 100 seconds of 1.024-s resolution data, prior to
burst trigger time, concatenated with at least 230 s of 64-ms resolution
data in order to fit a quadratic form to the background.  Similar results
were obtained with a first order polynomial background fit, but a quadratic
affords a flat burst region over at least 128-s in the vast majority of
cases.

For each subset of GRBs we aligned the highest peak, determined on a
1.024-s time scale, and cumulatively added $H$ for each time bin relative to
this highest peak.  Figure 1 depicts the difference in average hardness
profiles between the two burst groups.  For the bright burst sample, the
uncertainties are determined primarily by the differences among hardness
ratios of GRBs, rather than by inaccuracy in the determinations for
individual GRBs.  For dim bursts, statistical fluctuations play a larger
but still minority role in the overall error budget.  The error bars shown
therefore depict one sigma sample errors: 68\% inclusion of individual
determinations, reduced by $\sqrt{N-1}$, where $N$ is the sample size.

The sense of the spectral difference we have found is that the dim bursts
are on average softer than the bright bursts, in the interval within 8-s of
the peak of the average profiles.  We have found this spectral difference
in several different ways using different subsets of the BATSE data. From
the error bars depicted in Figure 1, we see that the hardness ratios of the
bright and dim groups differ at about the 3 sigma level near the central
time bin, and at about the one sigma level for many of the other time bins
tested.

\bigskip

\cl{\bf 3. EFFECT EVIDENT IN SECOND BATSE CATALOG}

As of this writing 482 BATSE detected GRBs have entered the public domain
which have recorded fluences and peak fluxes (Meegan et al. 1994). To
search for gross spectral differences in the public domain data, we used
the fluences for channels 1, 2, 3 and 4 (referred to as $F_1$ through $F_4$
respectively) as well as the peak flux on the 64 millisecond scale
integrated over all 4 channels ($P_{64}$) to perform the described test.

We first divided the data into a bright half and dim half as determined by
$P_{64}$.  Then, to test for a brightness - hardness correlation, we
defined the same type of cumulative hardness sum used in equation (1)
($H_{i/j}$ between energy bands $i$ and $j$, except this time using the
whole fluence of bursts instead of flux at given relative times) for both
samples and compared them statistically.  We searched for a brightness -
hardness correlation using channels 4 and 3, channels 3 and 2, and channels
2 and 1 for each sample. One $\sigma$ errors were determined as above from
the sample variance.

Values for these hardness ratios are shown in Table 1.  Column 1 of this
table gives the energy channels being used.  For example, a ``$3/2$" in
column 1 denotes a row where the hardness ratio of the fluence in channel 3
divided by the fluence in channel 2 is given. Column 2 lists the hardness
statistic for the first 2 years of BATSE data in the public domain with
$P_{64}$ below the median value.  Column 3 lists the same hardness
statistic for these GRBs in the public domain with $P_{64}$ above the
median value.  Column 4 lists the significance of the difference between
the column 2 and 3 values as determined from the combined variances of both
samples.  This significance implicitly assumes a Gaussian form for the
variance.

{}From inspection of Table 1 we see that the high $P_{64}$ half of the public
domain bursts are significantly harder than the dimmer half.  High
statistical significance, above the 4 $\sigma$ level, is seen both when
computing the cumulative hardness ratios between channels 3 and 2, and
between channels 2 and 1. The low statistical significance between channels
4 and 3 is primarily due to the low number of GRBs with significant fluence
in channel 4.

Figure 2 shows graphically the results of a similar test done with more
divisions of $P_{64}$.  Instead of breaking the samples up into only 2
groups of high and low peak flux, we broke the public domain sample up into
5 groups.  The plot shows that the gross relation between peak flux and
hardness is not dominated by particularly bright bursts or by weak bursts
near the BATSE sensitivity limit.  The brightness-hardness relation is
discernable over the whole data sample.  We feel this result complements
the conclusion reached in the previous section.

\bigskip

\cl{\bf 4. POSSIBLE SYSTEMATIC ERRORS AND BIASES }

The gross spectral difference between the bright and dim GRBs that we
have measured could be a property of GRBs themselves.  However, it
could also be due to systematic errors or biases.  Sources of these
include biases introduced by BATSE detection thresholds and trigger
criteria, statistical fluctuations, assumptions applied to the data
analysis, or a combination of these effects.  A more detailed
discussion of some these possibilities follows.

One concern with a statistic that involves a ratio is that the results are
dominated by a few data points where the denominator is near zero (Laros et
al. 1984; Schaefer 1993).  We have addressed this problem by incorporating
the numerator as an additive term in the denominator; this measure
results in each contributing term in Eq. (1) being near unity.  Next, we
have exclusively used statistics that involve a cumulative sum.  None of
the denominators used in the \S 2 results were within 2 $\sigma$ of zero.
Thus, we are relatively immune to potential singularities that occur in
hardness ratio calculations.

Another concern involves statistical fluctuations in the measurements. An
indication that these fluctuations are not important comes from tracking
the accumulation of this measurement error (in quadrature) for our public
domain results.  Here, we have found that our results are extremely
significant, usually to better than 25 $\sigma$, when measurement error
{\it only} is examined.  This error is hence insignificant when compared to
fluctuations caused by the differences in spectral properties between
sample GRBs themselves.

The hardness difference could also be caused by the fluke inclusion of a
small sample of particularly bright, hard bursts or particularly dim, soft
bursts which carry unusually high statistical weight. Our choice of $H$ in
Eq. 1 was made to minimize this effect.  Each term in the Eq. (1) sum is
of order unity and has roughly equal statistical weight.  Thus, bursts near
the bright edge of a brightness group do not dominate the statistics.
Figure 2 also shows that the hardness difference is seen over the whole of
the public domain sample. To further test for sample variations, the
results were recalculated using randomly chosen subsets of the data.  These
tests did not indicate that such a fluke inclusion effect was operating.

Another concern is that we are testing for a correlation between quantities
that are not fully independent.  Perhaps $P_{64}$, whose counts are
effected mostly by channel 3, is intrinsically correlated with $H_{3/2}$,
whose value might contain a residual proportionality  to the counts in
channel 3. We argue that the correlation is seen in $H_{4/3}$, $H_{3/2}$
and $H_{2/1}$.  Were peak count rate dominated by a single channel (channel
3 for example), one might expect the opposite correlation between $P_{64}$
and $H_{4/3}$ than between peak count rate and $H_{3/2}$. Such an effect is
not seen.

The trigger algorithm of the BATSE instrument may introduce a bias into the
sample it detected, in which case our result would not be a measure of GRB
properties, but instead related to the trigger criteria.  BATSE triggers
(begins a GRB data accumulation) when the combined counts in channels 2 and
3 for two detectors on either the 64 ms, 256 ms, or 1024 ms time scale are
found to exceed 5.5 $ \sigma$ above the background determined from the
previous 17 s (Fishman et al. 1989).

One example of such a ``trigger bias" for weak bursts is that, statistical
fluctuations causing detection may favor GRBs with $H_{3/2}$ near unity and
bias the sample toward this value.  However, this is harder than the
measured results in Figure 1 and could not further soften the dim burst
sample.  In another example, the trigger criteria demands that counts in
channels 2 and/or 3 (which typically have the highest count rates) be above
a certain level, but allows arbitrarily low counts in channels 1 and 4.
This could introduce a bias against bursts with high counts in 1 and 4 and
low counts in 2 and 3 causing a systematic softening in the measured
$H_{4/3}$ for dim bursts.  However, if this bias were significant it would
also cause $H_{2/1}$ to harden, and both effects are not consistent with
the statistical softening of bursts detected.

We have tested for several other systematic errors including potential
background effects and selection criteria without finding an effect which
could successfully mimic the proposed dim/soft correlation claimed. In
conclusion, we believe that bias is not responsible for the measured
difference in spectral hardness ratios. We therefore assert that the
measured gross spectral differences of GRBs originate from properties of
the GRBs themselves.

\bigskip

\cl{\bf 5. DISCUSSION }

Could this brightness - hardness correlation be related to bulk special
relativistic motion?  Perhaps the brighter bursts are being beamed toward
us at a significantly higher Lorentz factor. One might then also expect a
correlation between duration and spectral hardness - and such a correlation
has been reported recently by Kouveliotou et al. (1993). Were SR effects to
completely explain the different durations of GRBs, the shortest GRBs would
be expected to be seen 3 decades in energy higher than the longest duration
GRBs. At first glance, one might conclude that such an effect is not seen -
the Kouveliotou et al. (1993) effect is more modest.  However, to rule this
out definitively one must run Monte Carlo calculations testing how these
great spectral shifts combine with BATSE's trigger criteria.

Could this brightness - hardness correlation be related to cosmology? A
redshift of order unity is expected from the number - brightness relation
as analyzed by Wickramasinghe et al. (1993), by Fenimore et al. (1993), and
by Emslie and Horack (1994). A time-dilation effect of order a factor of
two between bright and the dim burst groups has recently been measured by
Norris et al. (1993a, 1993b, 1994).  That more distant GRBs would have
their spectra redshifted relative to nearby GRBs is in the same sense as
the observed result.  It is therefore possible that the observed hardness -
brightness correlation is primarily a result of cosmological expansion of
the universe.

Relativistic effects, however, might work in the opposite direction than
that proposed above (Turner 1993). A soft GRB with a power-law spectra
would have its spectra redshifted an equal amount as a hard GRB, but more
energy would be shifted out of BATSE's trigger channels, which could cause
it to fall below BATSE's trigger detection threshold. This cosmological
effect would work in the opposite direction from what is observed - making
detected dim GRBs {\it harder}.

GRB spectra, however, are not well described by a power law - a fact which
underlies the observed difference in hardness ratios.  If GRB spectra fall
off more rapidly at higher energies, this rapid fall off would also be
shifted to the BATSE detection bands, and could mean that some dim GRBS
would be measured as {\it softer} in the shifted energy ranges.  To
determine the magnitude and even {\it direction} of this effect, one must
run Monte Carlo simulations.

These effects are primarily related to the triggering of GRBs, but, as seen
in Figure 2, the brightness-hardness relation is visible even for sets of
GRBs significantly above BATSE's triggering threshold.  Given our sample
completeness with respect to trigger criteria, this artificial threshold
should affect each group in the same way.

Some might object to our use of peak count rate as a discriminative
attribute of GRBs rather than fluence. For all hardness ratio formulations
involving fluence, however, the results were statistically insignificant.
This suggests that peak flux is a more discriminative attribute than
fluence for GRBs (see also Kouveliotou et al. 1993).

In summary, we have found that within the sample of BATSE-detected bursts,
bright GRBs are spectrally harder than dim ones.  We believe that this is
the first time that fluctuations in the mean hardness caused by the
differing properties of bursts in the sample have been specifically
addressed in determining the significance of the difference in hardness in
GRB brightness subclassses. The cause of the spectral difference is open to
question, but it appears that it may be related to the time dilation
measured in Norris et al. (1994).  We are currently working to determine
whether a time dilation consistent with Norris et al. (1994) adequately
describes this measured spectral correlation (Bonnell et al. 1994). Our
current best guess is that this correlation is cosmological in origin.

\bigskip

We thank Bohdan Paczynski and Bradley Schaefer for helpful comments and
discussions.  We are particularly grateful to David Palmer for a
independent, qualitative check of our results on public domain data. This
research was supported by a grant from NASA.

\vfill\eject

\def\tabletoprule{\noalign{\hrule\smallskip}}
\def\tablerule{\noalign{\smallskip\hrule\smallskip}}
\def\tablebottomrule{\noalign{\smallskip\hrule}}

\ \ \ \ \ \ \ \ \ \ \ \ \ \ \ \ \ \ \ \ \ \ \ \ \ \ Table 1

\noindent
\ \ \ \ Hardness Ratios of First 2 Years of BATSE GRBs

\bigskip

\halign{#\hfil& \quad\hfil#& \quad\hfil#& \quad\hfil#& \quad\hfil#\cr
\tabletoprule
\tabletoprule
Hardness & $H$ for GRBs & $H$ for GRBs & $\sigma$~~~~~~   \cr
Channels & $P_{64}$ $<$ median & $P_{64}$ $>$ median & difference \cr
\tablerule
{}~~~~$2/1$&   1.34~~~~~~   &  1.68~~~~~~~ &   5.2~~~~ \cr
{}~~~~$3/2$&   2.21~~~~~~   &  3.25~~~~~~~ &   6.8~~~~ \cr
{}~~~~$4/3$&   1.13~~~~~~   &  1.39~~~~~~~ &   1.5~~~~ \cr
\tablebottomrule}

\vfill\eject

\cl{\bf REFERENCES }
{
\parindent=0pt
\hangindent=20pt
\baselineskip=18pt
\parskip=4pt

\bigskip

\hangindent=20pt
Atteia, J.L., Barat, C., Hurley, K., Niel, M., Vedrenne, G., Evans, W.D.,
Fenimore, E.E., Klebesadel, R.W., Laros, J.G., CLine, T., Desai, U.,
Teegarden, B., Estulin, I.V., Zenchenko, V.M., Kuznetsov, A.V., and Kurt,
V.G. 1987, ApJ Sup. 64, 305

Band, D. L. et al., 1993, ApJ, 413, 281

Bonnell, J. T. et al. 1994, in preparation

\hangindent=20pt
Davis, S.P., Norris, J.P., Kouveliotou, C., Fishman, G.J., Meegan, C.A., \&
Paciesas, W.S.  1994, in 1993 Huntsville Gamma-Ray Burst Workshop, eds: G.
Fishman, K. Hurley, and J. Brainerd (New York: AIP), in press

\hangindent=20pt
Dezalay, J.-P., Barat, C., Talon, R., Sunyaev, R., Terekhov, O., and
Kuznetsov, A. 1992, in Gamma Ray Bursts, AIP Conference Proceedings
265, eds: W.S. Paciesas \& G.J. Fishman, (New York: AIP), 304

Emslie, G. A. and Horack, J. M. 1994, ApJ, in press

Fenimore, E.E., et al.  1993, Nature, in press

\hangindent=20pt
Fishman, G.J., et al.  1989, in Proc. Gamma Ray Observatory Science
Workshop, ed W.N. Johnson (Greenbelt, MD: NASA/GSFC), p 2-39

\hangindent=20pt
Fishman, G.J. et al. 1992, in Gamma Ray Bursts, AIP Conference Proceedings
265, eds: W.S. Paciesas \& G.J. Fishman, (New York: AIP), 13

\hangindent=20pt
Kouveliotou, C., Meegan, C.A., Fishman, G.J., Bhat, N.P., Briggs, M.S.,
Koshut, T.M., Paciesas, W. S. and Pendleton, G. N. 1993, ApJ, 413, L101

Laros, G.J. et al. 1984, ApJ, 286, 681

\hangindent=20pt
Meegan, C.A., Fishman, G.J., Wilson, R.B., Paciesas, W.S., Pendleton, G.N,
Horack, J.M., 1992, Nature, 355, 143

\hangindent=20pt
Meegan, C.A., Fishman, G.J., Horack, J.M., Brock, M.N., Cole, S. Paciesas,
W.S., Briggs, M.S. Pendleton, G.N. Preece, R., Koshut, T.M., Mallozzi
R.S., Kouveliotou, C. and McCollough, M. 1994, public domain data
available at grossc.gsfc.nasa.gov with username GRONEWS

Mitrofanov, I. et al. 1992a, Sov. Astron. J., 69, 1052

\hangindent=20pt
Mitrofanov, I. et al. 1992b, in Gamma Ray Bursts, AIP Conference
Proceedings 265, eds: W. S. Paciesas \& G. J. Fishman (New York, AIP), 163

\hangindent=20pt
Mitrofanov et al. 1993, in Compton Gamma Ray Observatory, AIP Conference
Proceedings 280, eds: M. Friedlander, N. Gehrels, \& D. Macomb (New York:
AIP), 761

\hangindent=20pt
Mitrofanov, I.G., Chernenko, A.M., Pozanenko, A.S. Paciesas, W.S.,
Kouveliotou, C., Meegan, C.A., Fishman, G.J., and Sagdeev, R.Z. 1994, in
1993 Huntsville Gamma-Ray Burst Workshop, eds: G. Fishman, K. Hurley, and
J. Brainerd (New York: AIP), in press

Nemiroff, R.J., 1994, Comments Ap, 17, 189

\hangindent=20pt
Norris, J.P., Davis, S.P., Kouveliotou, C., Fishman, G.J., Meegan, C.A.,
Wilson, R.B., \& Paciesas, W.S.  1993a, in Proc. Compton Symp., ed. M.
Friedlander, N. Gehrels, \& D. Macomb (New York:  AIP), 959

\hangindent=20pt
Norris, J.P., Nemiroff, R.J., Kouveliotou, C., Fishman, G.J., Meegan, C.A.,
Wilson, R.B. \& Paciesas, W.S.  1993b, in Proc. Compton Symp., ed. M.
Friedlander, N. Gehrels, \& D. Macomb (New York:  AIP), 947

\hangindent=20pt
Norris, J. P., Nemiroff, R. J., Scargle, J. D., Kouveliotou, C., Fishman G.
J., Meegan, C. A., Paciesas. W. S., and Bonnell J. T. 1994, ApJ, 424, 540

\hangindent=20pt
Paciesas, W. S., Pendleton, G. N., Kouveliotou, C., Fishman, G. J., Meegan,
C. A., and Wilson, R. B. 1992, in Gamma Ray Bursts, AIP Conference
Proceedings 265, eds: W. S. Paciesas \& G. J. Fishman (New York, AIP), 190

Paczynski, B.  1992, Nature, 355, 521

Piran, T.  1992, ApJ, 389, L45

Schaefer, B. E. et al. 1992, ApJ 393, L51

Schaefer, B. E. 1993, ApJ, 404 L87

Schaefer, B. E. et al. 1994, ApJ Supp., 92, 285

Turner, E. 1993, private communication to B. Paczynski

Usov, V. V. \& Chibisov G. V. 1975, Sov. Astron. J., 19, 115

van den Berg, S. 1983, Ap\& SS, 97, 385

\hangindent=20pt
Wickramasinghe, W.A.D.T., Nemiroff, R.J., Norris, J.P., Kouveliotou, C.,
Fishman, G.J., Meegan, C.A., Wilson, R.B., \& Paciesas, W.S. 1993, ApJ,
411,
L55

}

\vfill\eject

\cl{\bf FIGURE CAPTIONS }
\hangindent=0pt
\baselineskip=18pt
\parindent=0pt

{\bf Figure 1:}  The boxes depict hardness ratio (channel 3
divided by channel 2) versus time for the bright burst group, while the X's
denote hardness versus time for the dim burst group. One sigma errors bars
determined from the sample variance are depicted by the crosses.

\bigskip

{\bf Figure 2:} A plot of hardness ratio in fluence versus peak flux for
the first two years of BATSE data (which includes 482 public domain GRBs).
Peak flux is taken on the 64-ms time scale.  The sample has been broken up
into 5 brightness groups. One sigma error bars, determined from the sample
variances internal to each brightness group, are shown.

\vfill\eject
\end